\documentclass[aps,prl,twocolumn]{revtex4-1}
\usepackage{graphicx}
\usepackage{amsmath}
\usepackage{amssymb}
\usepackage{url}
\usepackage[caption=false]{subfig}
\usepackage{units}
\usepackage{color}
\usepackage[normalem]{ulem}

\begin{document}
\title{Attosecond probing of nuclear dynamics with trajectory-resolved high-harmonic spectroscopy}
\author{Pengfei Lan$^1$, Marc Ruhmann$^2$, Lixin He$^1$, Chunyang Zhai$^1$, Feng Wang$^1$, Xiaosong Zhu$^1$, Qingbin Zhang$^1$, Yueming Zhou$^1$, Min Li$^1$, Manfred Lein$^{2}$}
\email{Corresponding author: lein@itp.uni-hannover.de }
\author {Peixiang Lu$^{1,3}$ }
\email{Corresponding author: lupeixiang@mail.hust.edu.cn}
%
\affiliation{$^1$ School of Physics and Wuhan National Laboratory for Optoelectronics, Huazhong University
of Science and Technology, Wuhan 430074, China \\
$^2$Institute for Theoretical Physics and Centre for Quantum Engineering and Space-Time Research (QUEST), Leibniz Universit\"at Hannover, Appelstra{\ss}e 2, 30167 Hannover, Germany\\
$^3$Laboratory of Optical Information Technology, Wuhan Institute of Technology, Wuhan 430205, China}

\begin{abstract}
We report attosecond-scale probing of the laser-induced
dynamics in molecules.
We apply the method of high-harmonic spectroscopy, where laser-driven recolliding electrons
on various trajectories record the motion of their parent ion.
Based on the transient phase-matching mechanism of high-order harmonic generation,
short and long trajectories contributing to the same harmonic order are distinguishable in both the spatial and frequency domains, giving rise to a one-to-one map between time and photon energy for each trajectory. The short and long trajectories
{in H$_2$ and D$_2$}
are used simultaneously to retrieve the nuclear dynamics on the attosecond and {\aa}ngstr\"{o}m scale.
Compared to using only short trajectories, this
extends the temporal range of the measurement to one optical cycle.
{The experiment is also applied to  methane and ammonia molecules.}

\end{abstract}

\maketitle

There is a continuous desire to develop methods with ever-better resolution in ultrafast science. Generally, time-resolved methods rely on ultrashort laser pulses. Therefore in the past decades, substantial efforts have been paid to produce attosecond pulses based on high-order harmonic generation (HHG) \cite{atto1,pump1,atto2,atto3}.
Since the duration of attosecond pulses is comparable to the time scale of
bound electrons, such sources
provide an important tool for detecting the ultrafast electron dynamics inside atoms
or molecules \cite{dyna1,dyna2,md,pump2}, inaugurating a new domain for time-resolved metrology and spectroscopy \cite{fk}.

One popular method of attosecond probing is the pump-probe measurement \cite{pump1,pump2}.
In this scheme, a physical process is first triggered by an attosecond pump pulse and subsequently probed by a near-infrared pulse (or vice versa). Then the time-dependent information can be decoded from the streaked photoelectron spectra recorded at different pump-probe delays.
On the other hand, HHG itself is a sub-femtosecond nonlinear process arising from laser induced electron-ion recollisions \cite{3step}.
Rich information about the electron-ion system at the time of recollision is encoded in the harmonic spectra. Extracting this information systematically is known as high-harmonic spectroscopy (HHS).
HHS has been exploited to image molecular structure with {\aa}ngstr\"{o}m resolution
\cite{it,ps,Vozzi2011,str1,str2,str3}, e.g., by molecular orbital tomography \cite{it,ps,Vozzi2011}.
Moreover, for each
high-harmonic order,
the freed electron spends a specific time in the continuum before recollision, resulting in the temporal chirp of  \mbox{HHG} \cite{mairesse}.
The ionization-recollision delay is analogous to a pump-probe delay,
providing an alternative way to map the photon energy to time.
Based on this property, HHS has been developed into an emerging tool for studying nuclear dynamics \cite{lein,baker,patch,farrell,foerster,krauswoerner} and charge migration \cite{worner,d2,dy3,dy4} with attosecond temporal resolution. However, according to the recollision model \cite{3step}, there are two trajectories per optical half cycle contributing to each individual harmonic order. The two trajectories are referred to as the ``short" and ``long" trajectories. 
Therefore, to guarantee a one-to-one map between the photon energy and time, one trajectory has to be selected by adjusting the phase-matching of HHG. For instance, in \cite{baker}, only the short one is relevant. Since HHG from the long trajectory shows a different chirp property \cite{mairesse} compared to the short trajectory, HHS will become more powerful if both can be utilized simultaneously.

%

In this Letter, we demonstrate a trajectory-resolved HHS method. Based on the time-dependent phase-matching of HHG, the short and long trajectories are successfully separated in both the spatial and frequency domains, which enables us to build the one-to-one map for each trajectory.
Then the short and long trajectories are simultaneously employed to retrieve the laser-induced proton dynamics of hydrogen molecules. This effectively extends the temporal range and enriches the information of the measurement.
On the theory side, the present work resorts to complex-time trajectories also known as quantum orbits \cite{sfa,salieres}. It thus goes beyond the previous approach \cite{lein,baker} based on
classical trajectories.
Therefore, our theory includes the physical effect that the ionization probability
depends on the internuclear distance $R$ so that the vibrational wave packet launched
by strong-field ionization differs from the vibrational ground state of the neutral molecule
\cite{urbain}. Compared to \cite{foerster,krauswoerner}, where this effect was included
by using $R$-dependent or vibrational-level dependent tunneling ionization rates, the
complex-time approach is not limited to the tunneling (low-frequency) regime.

\begin{figure}[t]
\includegraphics[width=\columnwidth]{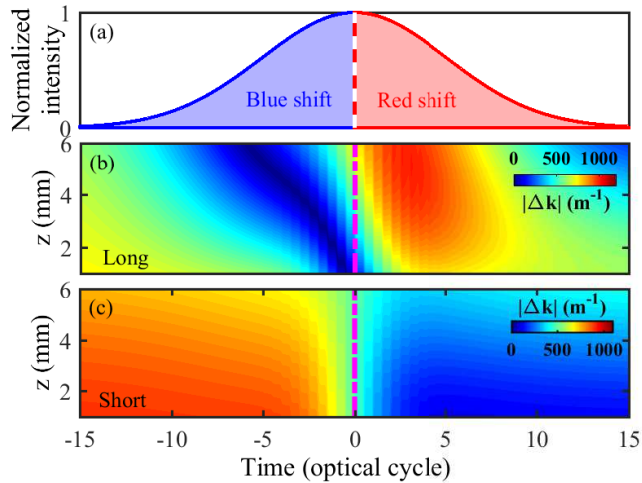}
\caption{\label{fig:phasematch}(a) Normalized envelope of the laser pulse. (b) and (c) Time-dependent phase mismatch of 17th harmonic for the long and short trajectories. {In the simulation, a 30-fs, 800-nm laser field with intensity 1.5 $\times10^{14}\,{\rm W/cm^2}$ is adopted}.}
\end{figure}

An important ingredient of our scheme is provided by transient phase-matching. To explain the
idea, let us revisit the Lewenstein model of HHG \cite{sfa,salieres}: HHG can be
described by the coherent sum over all different quantum trajectories. Each trajectory
carries a dipole phase $\varphi^j_q$. It can be approximately written as $\varphi^j_q\simeq\alpha^j_qI(t)$, where $I(t)$ is the driving laser
intensity and $\alpha^j_q$ is the coefficient of $q$-th harmonic with $j = S$ or $j= L$ representing the
short and long trajectories, respectively.
In a short laser pulse, the intensity varies with time so that the
dipole phase $\varphi^j_q$ is time dependent. This leads to an instantaneous frequency of each harmonic, which can
be expressed by \cite{heyl,he,feng,mbg}
$\omega_q=q\omega_0+\alpha^j_q\frac{\partial I(t)}{\partial t}$.
Hence, the  harmonic emission {in} the leading ($\frac{\partial I(t)}{\partial t}>0$)/falling ($\frac{\partial I(t)}{\partial t}<0$) edge of a pulse implies a blue/red shift in the harmonic spectrum
[{see} Fig. \ref{fig:phasematch}(a)].
On the other hand, the HHG process involves macroscopic
propagation effects in the gas medium. {The phase mismatch
{is} $\Delta k = \Delta k_g + \Delta k_d + \Delta k_e + \Delta k_a$ \cite{feng,mbg}. Here $\Delta k_g=q\frac{2}{b[1 + (2z/b)^2 ]}$ is due to the spatial phase variations arising from the focusing geometry of the driving laser, $q$ is the harmonic order, $b$ is the confocal parameter, $z$ is the position of the medium. $\Delta k_d = \alpha_j \frac{dI} {dz}$
is the phase mismatch due to the
intensity-dependent dipole phase. $\Delta k_e = \frac{\omega}{c} \Delta n_{el} (\omega)$
and $\Delta k_a = \frac{\omega}{c} \Delta n_{at} (\omega)$ are due to the dispersions of the free
electrons and neutrals. $\Delta n(\omega)$ is the difference between the refractive
indices at the harmonic and at the driving laser frequency}. It depends on the densities of free electrons and neutrals.
The time-dependent intensity of an ultrashort pulse leads to variations of the dipole phase and ionization probability
and thus to time-dependent $\Delta k_d$, $\Delta k_e$, $\Delta k_a$ and $\Delta k$.
In Figs. \ref{fig:phasematch}(b),(c), we show the values of $|\Delta k|$ of the 17th harmonic for the long and short trajectories driven by a 30-fs, 800-nm laser field with intensity 1.5 $\times10^{14}\,{\rm W/cm^2}$.
The long trajectory is better phase matched on the leading edge, while the short trajectory is favored on the falling edge. Therefore, {HHG from the} long trajectory will be blue shifted and that of the short trajectory will be red shifted. This enables us to clearly separate these two trajectories in the frequency domain. Moreover, since the coefficient $\alpha^L_q$ of the long trajectory is larger than $\alpha^S_q$ of the short trajectory, the phase-matching angle for the long trajectory is larger. {Thus the short and long trajectories can also be distinguished in the spatial domain.} In short, it is possible to separate the short and long trajectories by recording the spatial profile and spectral properties of high harmonic spectra.
The trajectory-resolved HHS enables us to build the one-to-one map between the photon energy and time for both the short and long trajectories, which can be effectively used for ultrafast measurements using the method {of} \cite{lein,baker}.

\begin{figure}[t]
\includegraphics[width=\columnwidth]{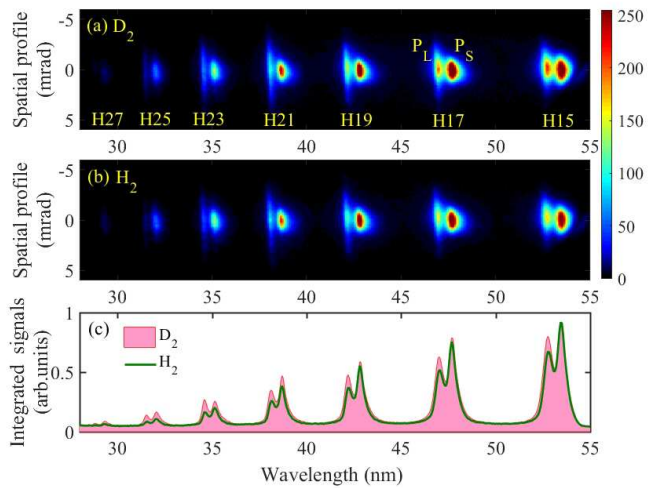}
\caption{\label{fig:hhgspectra}Spatially resolved  harmonic spectra from (a) D$_2$ and (b) H$_2$. (c) Spatially integrated
HHG signals for the spectra in (a) and (b).}
\end{figure}

A Ti-sapphire driving laser with a central wavelength of 800 nm and pulse duration of 30 fs is employed in our experiment.
The incident laser
beam is focused to a 2-mm-long gas cell with a pressure of 20 torr by a 600-mm focal-length lens. To evaluate the gas density in the gas cell, we measure the gas pressure with a species-independent vacuum gauge. As in \cite{he}, the phase matching of different trajectories is adjusted by carefully changing the beam size, laser power and gas position. Figures \ref{fig:hhgspectra}(a),(b) show the typical spatially resolved high harmonic spectra for D$_2$ and H$_2$. The laser intensity is estimated to be 1.5 $\times10^{14}\,{\rm W/cm^2}$.
Each harmonic is split into two peaks, marked as P$_\textmd{L}$ and P$_\textmd{S}$ for the 17th harmonic in Fig. \ref{fig:hhgspectra}(a). Moreover, the spatial profile of P$_\textmd{L}$ shows a larger divergence angle than that of P$_\textmd{S}$.
Following the above discussion, P$_\textmd{L}$ is due to the long trajectory and P$_\textmd{S}$ is due to the short trajectory.
Figure \ref{fig:hhgspectra}(c) shows the spatially integrated signals for the spectra of D$_2$ (full-filled) and H$_2$ (green solid line).
The short and long trajectories appear as separated peaks.
Unlike \cite{Zair},
our setup is adjusted to separate the trajectories and no obvious
interference between them is observed.

To analyze {macroscopic effects}, we have measured HHG from H$_2$ and D$_2$ as a function of the gas pressure \cite{corkum,wei}.
With the pressure changing from 15 to 35 Torr, the HHG yields of both
trajectories exhibit a quadratic increase (The scaling factors are 1.92, 2.05, respectively).
{The quadratic increase indicates
that reabsorption and phase mismatch due to ionization are insignificant in our experimental
conditions \cite{corkum,wei}.}

\begin{figure}[t]
    \subfloat{\includegraphics[width=.5\columnwidth]{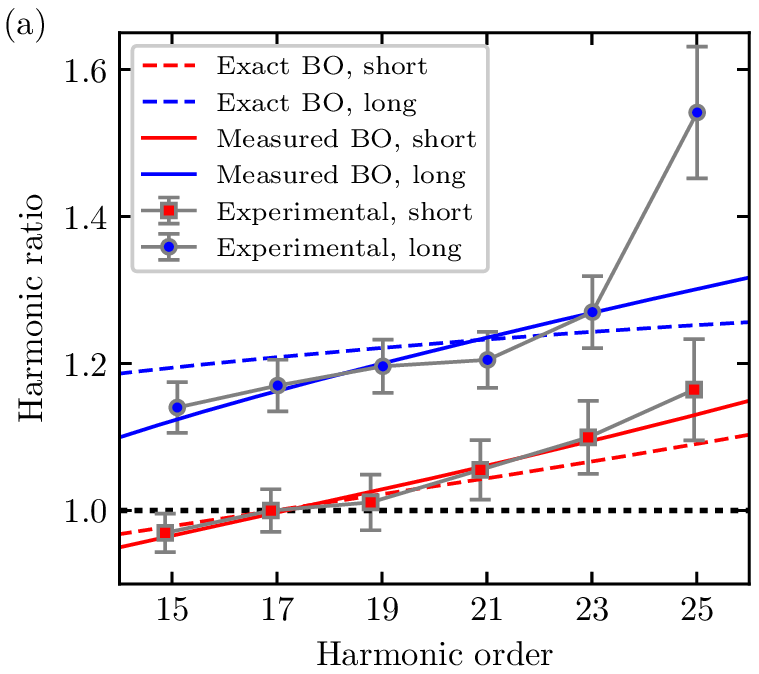}}
    \subfloat{\includegraphics[width=.5\columnwidth]{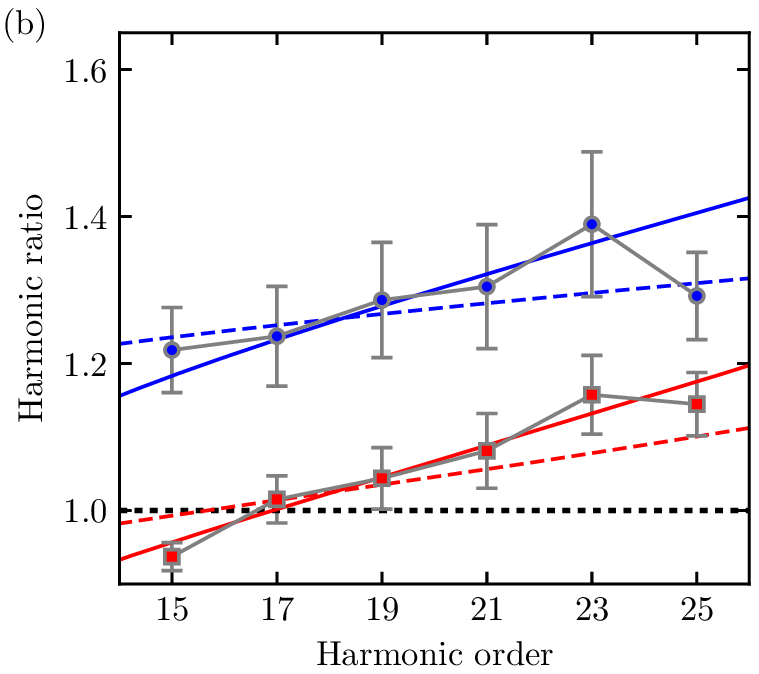}}
\caption{Ratio of harmonic intensities between D$_2$ and H$_2$ as a function of harmonic order. Circles and squares are for the long (peak P$_\textmd{L}$) and short (peak P$_\textmd{S}$) trajectories, respectively. The solid and dashed lines are the ratios $R_C(\omega$), calculated with complex saddle-point times (see text), for the measured and exact field-free BO potentials, respectively. The laser intensity is (a) $1.5\times10^{14}$ and (b) $2\times10^{14}$ W/cm$^2$.}
\label{fig:dhratio}
\end{figure}


Figure \ref{fig:hhgspectra}(c) shows that the harmonic signal of D$_2$ is stronger than that of H$_2$. This can be more clearly seen from Fig. \ref{fig:dhratio}(a), which shows the harmonic ratio between D$_2$ and H$_2$ as a function of harmonic order.
The ratio is calculated by using the peak values
in the spatially integrated spectra.
Gaussian fitting to the peaks and integrating the area under each peak leads to similar results within the error bars in Fig. \ref{fig:dhratio}(a).
Harmonics generated from D$_2$ are {mostly} stronger than those from H$_2$, thus the ratio is larger than 1. Moreover, the ratio increases with the harmonic order.
The ratio of long trajectory is larger than that of short trajectory.
Similar results are obtained for a higher laser intensity of $2\times10^{14}$ W/cm$^2$, see Fig. \ref{fig:dhratio}(b).
The increasing ratio for the short trajectory is consistent with the earlier results reported in \cite{lein,baker}.

To explain our experiment and
to retrieve the proton dynamics we use the theory developed in \cite{lein,baker,lein3} extended to long trajectories and complex-time electron trajectories.
The vibrational dynamics within the Born-Oppenheimer (BO) approximation can be readily included into Lewenstein model \cite{sfa} for HHG \cite{lein,baker,lein3}. When the neutral molecule is ionized, in addition to the continuum motion of the electron, vibrational dynamics is initiated in the remaining parent
ion,
leading to
the appearance of the vibrational autocorrelation function.
Including the dependence of the recombination transition matrix element $d_\mathrm{rec}$ on the internuclear distance $R$, the harmonic intensity for an electron trajectory with ionization time $t'$ and recombination time $t$
is proportional to the squared modulus of \cite{lein1}
\begin{align}
	C(t,t') = \int\limits_0^\infty \!\mathrm{d}R \, d_\mathrm{rec}^*(\mathbf{k}_\mathrm{r}(t,t'),R) \chi_0^*(R) U_R^+(t\!-\!t') \chi_0(R).
	\label{eqn:autocorrelation}
\end{align}
Here, $\chi_0$ is the vibrational ground state of the neutral molecule and $\mathbf{k}_\mathrm{r}(t,t')$ is the return momentum of the recolliding electron.
The time-evolution operator $U_R^+$ describes the vibrational motion on the BO potential curve of the ion H$_2^+$/D$_2^+$. According to (\ref{eqn:autocorrelation}), the probability for electron recombination is sensitive to the overlap between the evolved vibrational state of the ion and the ground vibrational state of the neutral molecule. The recombination matrix element $d_\mathrm{rec}(\mathbf{k},R) = \mathbf{e}_E \langle \mathbf{k} | \mathbf{\hat{p}} | \psi_R \rangle$ in the direction of the laser polarization $\mathbf{e}_E$ is taken in velocity form \cite{lein1}. It accounts for two-center interference effects in the recombination step \cite{lein4, lein5, vozzi, kanai}. The continuum states $|\mathbf{k}\rangle$ are approximated by plane waves $e^{\mathrm{i} \mathbf{k}\cdot\mathbf{r}}$ and $\psi_R$ is the electronic Dyson orbital for fixed $R$. The latter can be approximated by a linear combination of hydrogen ground states \cite{lein6}, yielding $d_\mathrm{rec}(k,R) \propto \cos(kR\cos\theta/2)$, where $\theta$ is the angle between the momentum of the returning electron and the internuclear axis.

The times for ionization and recollision are obtained by applying the saddle-point approximation to the HHG amplitude. For a linearly polarized laser field, the relevant equations for harmonic frequency $\omega$ are \cite{salieres,phasehhg}
\begin{align}
	\frac{ \big[ p(t,t') + A(t') \big]^2 }{2} &= -I_\mathrm{p} \label{eqn:ion_time}\\
	\frac{ \big[ p(t,t') + A(t) \big]^2 }{2} &= \omega - I_\mathrm{p} \label{eqn:rec_time}
\end{align}
with the saddle-point momentum $p(t,t') = -\int_{t'}^t \mathrm{d}t'' A(t'') / (t - t')$.
Here, $A(t)=-\int^t \mathrm{d}t' \, E(t')$ is defined in terms of the
laser electric field $E(t)$. The ionization potential is defined as $I_\mathrm{p} = V_\mathrm{BO}^+(R_0) - E_0$ where $V_\mathrm{BO}^+$ is the BO potential of the ion. $R_0$ is the equilibrium distance of the neutral molecule and $E_0$ is its ground-state energy. We use the full complex solutions of (\ref{eqn:ion_time}) and (\ref{eqn:rec_time}) for $t'$ and $t$ in the calculations of the autocorrelation function. According to equation (\ref{eqn:rec_time}) the momentum of the electron at return is given by $k_\mathrm{r}(\omega) = \sqrt{2(\omega - I_\mathrm{p})}$.
The distribution $\sigma(\theta)$ of alignment angles $\theta$ is calculated at the pulse peak as in \cite{baker2008}, assuming room temperature 293K of the molecules. The summation over angles yields
\begin{align}
	C(\omega) = \int\limits_0^\infty \!\!\mathrm{d}R  \int\limits_0^{\pi/2} \!\!\mathrm{d}\theta\,\sigma(\theta) \cos\!\Big(\!{kR\over2}\cos\theta\!\Big) \,\chi_0^*(R)\, U_R^+\left(\tau\right) \chi_0(R)
\end{align}
with excursion time $\tau=\tau(\omega) = t(\omega)\!-\!t'(\omega)$. The short or long trajectory is selected by choosing the appropriate solutions of Eqs. (\ref{eqn:ion_time}), (\ref{eqn:rec_time}). The ratio of harmonic intensities from D$_2$ versus H$_2$ is then approximated as
\begin{align}
	R_C(\omega) = \frac{|C_\mathrm{D}(\omega)|^2 \, \Gamma\big(I_\mathrm{p}^\mathrm{D}, t_\mathrm{D}'(\omega)\big)}{|C_\mathrm{H}(\omega)|^2 \, \Gamma\big(I_\mathrm{p}^\mathrm{H}, t_\mathrm{H}'(\omega)\big)}
	\label{eqn:harmonic_ratio}
\end{align}
with the instantaneous ionization rate $\Gamma(I_\mathrm{p}, t') = \exp\!\big(-2(2I_\mathrm{p})^{3/2}/(3|E(\operatorname{Re}t')|)\big)$.

\begin{figure}[t]
	\captionsetup[subfloat]{farskip=0pt}
    \subfloat{\includegraphics[width=.85\columnwidth]{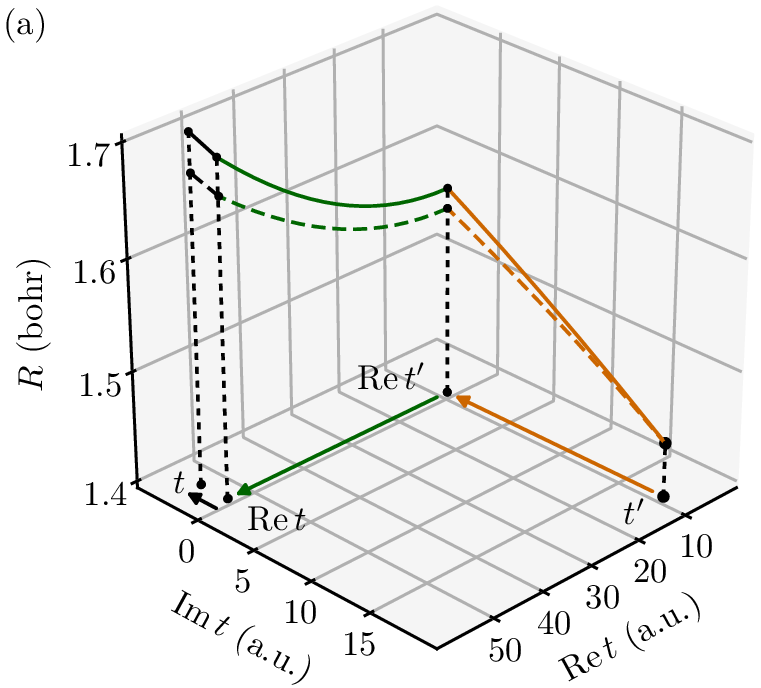}}

    \subfloat{\includegraphics[width=.85\columnwidth]{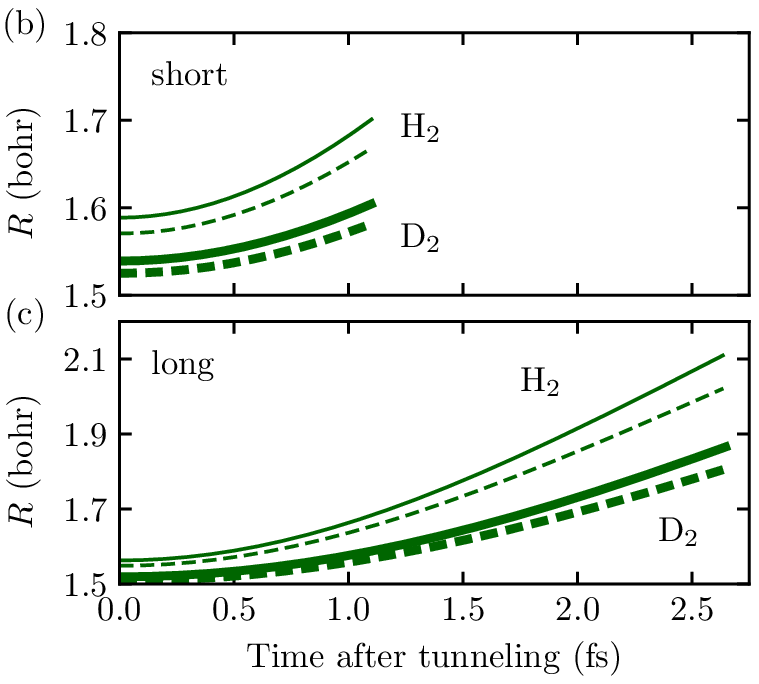}}
	\caption{Vibrational dynamics for the 19th harmonic in the exact field-free (dashed lines) and measured (solid lines) BO potentials. (a) Stepwise time evolution of the internuclear distance $R$ from the ionization time $t'$ to the recombination time $t$, for the short trajectory of H$_2$. (b) Evolution of the internuclear distance after tunneling, i.e. from $\operatorname{Re}t'$ to $\operatorname{Re}t$, for the short trajectory, for D$_2$ (thick lines) and H$_2$ (thin lines). The origin of the time axis is set to the moment after tunneling, i.e. to $\operatorname{Re}t'$. The curves start at different $R$ because the tunneling dynamics is dependent on the isotope and BO potential. (c) Same as (b) for the long trajectory.}
	\label{fig:nuclear_motion}
\end{figure}

The first section of the temporal path, namely $t' \rightarrow \operatorname{Re}{t'}$, indicated by the orange arrow in Fig. \ref{fig:nuclear_motion}(a), can be identified with the tunneling of the electron. During this time, the vibrational ground state of the neutral molecule undergoes an imaginary time evolution and yields the initial vibrational state of the ion for the subsequent real time evolution $\operatorname{Re}t' \rightarrow \operatorname{Re}t$. The last section  $\operatorname{Re}t \rightarrow t$ is usually small and does not affect the results significantly.

Previous theory \cite{lein,baker} has
already
shown that the increasing ratio for short trajectories is explained
by a slower decay of the autocorrelation function
in the heavier isotope due to the slower motion.
Since the dynamics of the cation is bound, it has a
turning point and already before reaching it, the autocorrelation
functions of the two isotopes approach each other, see Fig.~2
in \cite{lein1}. This explains that the harmonic ratio decreases
for long excursion times, namely in the limit of low harmonic orders
for long trajectories, cf. Fig. 2 in \cite{lein2}.
The location of the maximum ratio is also modified by
two-center interference
\cite{baker2008}, which is included in our present model.

We use an optimization algorithm \cite{mpfit} for finding an ionic BO potential such that the calculated ratios match the experimental ratios. This makes it possible to effectively measure the ionic potential in the range reachable within the time span of the short and long trajectories. {For this retrieval, we have neglected the small trajectory-dependent deviation of the harmonic frequencies from integer harmonic orders.}
The resulting ratios from this fit are shown in Fig. \ref{fig:dhratio} (a) and (b), reproducing the experimental ratios well both for intensities of $1.5$ and $2\times10^{14}$ W/cm$^2$. The corresponding nuclear motion for $1.5\times10^{14}$ W/cm$^2$ is depicted in Fig. \ref{fig:nuclear_motion}.
During the initial tunneling ($t' \rightarrow \operatorname{Re}t'$), the internuclear distance increases approximately linearly as Fig. \ref{fig:nuclear_motion}(a) shows.
Overall, the nuclei separate slightly more quickly than predicted by a calculation using the exact field-free BO potential.
Compared to the field-free BO potentials, the equilibrium
positions of the measured potentials are shifted to slightly larger
distance, e.g. from $R=2\,$a.u. to $R\approx 2.1\,$ a.u. ($R\approx2.25\,$a.u.) for the lower (higher intensity), and the energy of the potential
minimum is lowered by about 0.01$\,$a.u. (0.04$\,$a.u.).
Future work is required to assess
effects neglected in the present model, e.g. Stark effects \cite{Etches} or
time-dependent alignment distributions.

Figures \ref{fig:nuclear_motion}(b),(c) show that the long trajectories allow us to probe a considerably larger time range of the ionic vibrational dynamics, when compared to the short trajectory. Here, the origin of the time axis is set to the moment after tunneling. Since there is already vibrational motion taking place during the tunneling of the electron, the curves start at internuclear distances $R$ greater than the equilibrium distance of the neutral molecule of approximately \unit[1.4]{bohr}. This starting distance is dependent on the isotope and on the BO potential. Furthermore, the starting distance and the vibrational motion depend on the harmonic order. If classical times were used, i.e. if the dynamics of the nuclei during electron tunneling was not included, the curves in Figs. \ref{fig:nuclear_motion}(b),(c) would all start at the equilibrium distance of the neutral molecule. In this case the vibrational motion would be the same for all harmonic orders (but still dependent on the isotope).

\begin{figure}[t]
\includegraphics[width=\columnwidth]{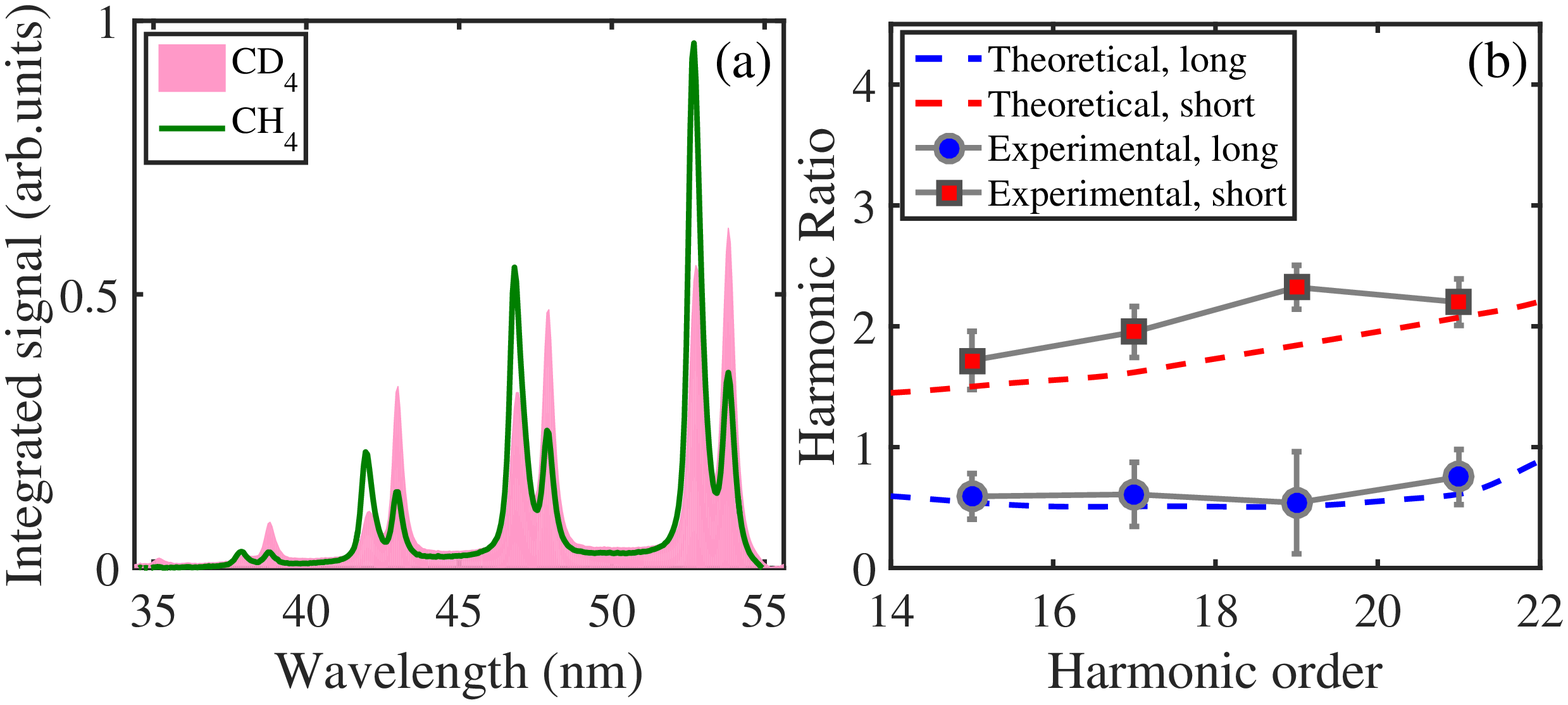}
\caption{(a) Spatially integrated high harmonic spectra from CH$_4$ and CD$_4$. (b) Ratio of harmonic intensities between CD$_4$ and CH$_4$ as a function of harmonic order. Circles and squares are for the long and short trajectories, respectively. The dashed lines show the theoretical ratio from \cite{ch4}.}
\label{fig:ch4}
\end{figure}

To explore the application of trajectory resolved HHS further,
we compare harmonic spectra obtained
from CH$_4$ and CD$_4$, see Fig. \ref{fig:ch4}.
The ratio of harmonic intensities between CD$_4$ and CH$_4$ is greater than 1 and increases with harmonic order for the short trajectory, which is in agreement with the previous experiment \cite{baker}. In contrast, the ratio for the long trajectory is less than 1 and also less than that of the short trajectory. This observed behavior agrees with the simulation result (dashed lines) in \cite{ch4} and is due to the changeover of the ratio {between the squared autocorrelation functions of CD$_4$ and CH$_4$ at 1.85 fs \cite{ch4,SM}.
Note that the previous experiment \cite{baker}, using only the short trajectory, did not cover the time range beyond 1.6 fs and hence did not exhibit ratios less than 1. We have also compared harmonics from NH$_3$ and ND$_3$ molecules. Good agreement between our experiment and simulation is found in
this case \cite{SM}.

In summary, we have confirmed that the isotope dependence of molecular high-order harmonic generation
persists for long trajectories. Using harmonic generation from both short and long trajectories
leads to an improved retrieval of the nuclear motion with attosecond and {\aa}ngstr\"{o}m precision. {We have demonstrated trajectory resolved HHS for  H$_2$/D$_2$, CH$_4$/CD$_4$, and NH$_3$/ND$_3$ molecules. This underlines the
general validity of the method.}

This work was supported by the National Natural Science Foundation of China under Grants
Nos. 11422435, 11234004 and 11404123.

\end{document}